\documentclass[11pt]{revtex4}
\usepackage{graphicx}
\usepackage{amsmath}

\begin{document}
\title{Sub-cycle waveform synthesis for effective control of electron localization }

\author{Pengfei Lan}
\email{pengfeilan@riken.jp}
\affiliation{Extreme Photonics Research Group, RIKEN Advanced Science Institute, 2-1 Hirosawa, Wako, Saitama 351-0198, Japan }
\author{Eiji J. Takahashi}
\email{ejtak@riken.jp}
\affiliation{Extreme Photonics Research Group, RIKEN Advanced Science Institute, 2-1 Hirosawa, Wako, Saitama 351-0198, Japan }
\author{Katsumi Midorikawa}
\affiliation{Extreme Photonics Research Group, RIKEN Advanced Science Institute, 2-1 Hirosawa, Wako, Saitama 351-0198, Japan }

\date{\today}

\begin{abstract}
We have proposed and demonstrated an effective scheme to steer both the
electronic and molecular motions of hydrogen molecule and its isotopes. Our scheme lies in the idea of sub-cycle waveform shaping (SWS) with
synthesizing coherent IR pulses of different wavelengths. The quantum model simulation indicates a remarkable enhancement of the electron localization asymmetry compared with the few-cycle pulse. $68\%$ and $89\%$ electrons of the dissociating fragmentation can be localized to one nucleus for $H_2^+$ and $H_2$, respectively.
\end{abstract}

\pacs{33.80.Rm, 33.80.Rv, 42.50.Hz, 42.65.Re}

\maketitle

Coherent control of quantum dynamics in chemical reactions and photoelectron processes
has been an exciting goal of chemistry and ultrafast physics. Such a process naturally involves both the electron and nuclear motions. On the one hand, femtosecond chemistries have made remarkable advance to control the pathway of molecular fragmentation using a optimally-shaped long pulse \citep{Zewail,Rabitz,Mies,Bandrauk,Cocke}.
On the other hand, the advent of carrier-envelope phase(CEP) stabilized few-cycle ($\sim5$ fs) pulse and the attosecond pulse have represented a robust tool to steer the electronic dynamics on the attosecond time scale.
One interesting question therefore is whether can we combine these two directions by effectively steering both the electronic and molecular motions.

It was demonstrated that, by using a few-cycle pulse, the electron wavepacket can be localized to one specific nucleus and therefore shows asymmetric electron localization for hydrogen molecules (or its isotopes) \citep{Esry,Kling,Ullrich} and the CO \citep{KlingCO}. Another approach to localize the electron was also proposed and demonstrated by synthesizing of fundament laser with another extreme ultraviolet \citep{He,CockeAPT,Sansone} or infrared pulse \citep{Ullrich2,Williams}. However, the localization asymmetry of the total dissociation in the few-cycle pulse is quite small \citep{Esry2007} even through the kinetic energy release (KER) spectrum reveal a large asymmetry. If the molecular wavepacket could respond to the electronic motion, it would be possible to control the electron location and chemical band in a very direct way.
Nevertheless, the precise steering of electronic motion requires the control of sub-cycle waveform, thus an extreme short pulse, e.g., 3.5-fs \citep{80as}, is desired.
On the contrary, the vibration period is longer than 14 fs even for the simplest molecule ion, $H_2^+$. The nuclear motion is too slow to respond directly to the electronic motion and the
few-cycle laser pulse. To meet this challenge, in this letter, we proposed and numerically demonstrated a novel scheme to effectively steer the
electronic and molecular motions of hydrogen molecule and its isotopes. Our scheme lies in the idea of sub-cycle waveform shaping (SWS) with
synthesizing coherent IR pulses of different wavelengths. We show that the SWS enables us to significantly enhance the asymmetry of electron localization compared with the few-cycle pulse.
\begin{figure}[!htp]
\vspace{1cm}
\begin{center}
\includegraphics[width=8cm]{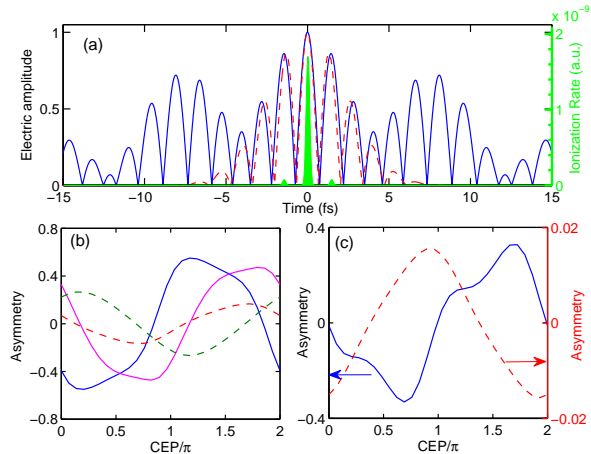}
\caption{\label{fig1}(Color)
(a) Temporal profiles of laser amplitude of the 4-fs fundamental (800 nm) OC pulse (red dashed line) and SWS pulse (blue line, 15 fs/800 nm + 25 fs/1200 nm).
The green filled line shows the ionization rate of $H_2$ subjected to the SWS field. The laser intensity is $1\times10^{14}$ W/cm$^2$ with an intensity ratio of $15\%$. $\phi_{\rm CE}$ and $\phi_1$ are 0.
(b) CEP dependence of the localization asymmetry. The solid magenta and blue lines correspond to the interaction of SWS pulse with 4th and 8th vibrational states of $H_2^+$,
respectively. The dashed red and green lines correspond to the interaction of 5-fs OC pulse with 4th and 8th vibrational states of $H_2^+$, respectively. (c) The CEP dependence of the Franck-Condon averaged localization asymmetry for the 5-fs OC field (red dashed line) and SWS pulse (solid blue line), respectively.
}
\end{center}
\end{figure}

With the development of laser technique,
it becomes possible to synthesize two coherent light sources output from different fiber laser \citep{Leitenstorfer}, optical parameter amplifier (OPA) \citep{EJtak,Oliver} or optical parameter chirp pulse amplifier (OPCPA) systems \citep{Kartner}. The basic idea of SWS stems from the temporal profile modulation by the beat wave of two IR waves \citep{Lan}.
By stabilizing the phase of the IR pulse, the waveform of the synthesis pulse can be precisely control. As shown Fig. \ref{fig1}(a), except the weak pedestal peaks, the synthesis
field with a 15-fs Ti:saphire laser and 25-fs IR (1200 nm) laser is equivalent to that of a 4-fs one-color (OC) field, which thus enables to precisely control the electronic motion. Moreover, the synthesized pulse contains pedestal peaks maintain a duration of about 15 fs, which is comparable to the vibration period of $H_2^+$ and nuclear motion can respond to these pedestal peaks. Therefore, we can expect an effective control of both the electronic and molecular motions with this SWS field.
We need to emphasize that to precisely control the electronic dynamics, it is important to shear the waveform of SWS pulse similar to a few-cycle OC pulse.
This condition requires that the pulse duration of the synthesized IR pulse should be less than 25 fs.

To demonstrate our scheme, we use a numerical
model that has been well established for studying electron localization with few-cycle pulse \citep{Kling,Ullrich,CockeAPT,Ullrich2,Williams}. This model solves the time-dependent Schr\"{o}dinger equation for the nuclear wavepacket in the Born-Oppenheimer (BO) representation. The time scale of the rotation is several hundreds fs, thus can be neglected in the ultrashort laser pulse. We concentrate on the hydrogen molecule and its isotopes, which has been of special interest for extensive investigation. The molecular dynamics can be well described in terms of the two lowest-lying electronic state. There are generally two kinds of experiments, one concentrates on the target of $H_2^+$ \citep{Esry,BenItzhak,Williams2000} and the other one on the $H_2$ \cite{Kling,CockeAPT}. The underlying physics is however very different. We first analysis the first situation and assume that $H_2^+$ is aligned along the direction of the laser field. Then the TDSE can be written as \cite{Kling}
\begin{equation}
i\frac{\partial}{\partial~t}\binom{\Psi_g}{\Psi_u}=-\dbinom{-\frac{\nabla^2}{(2\mu)}+V_g\qquad~ V_{gu}}{V_{gu}^*\qquad~-\frac{\nabla^2}{(2\mu)}+V_u}\dbinom{\Psi_g}{\Psi_u},
\end{equation}
where $\mu$ is the nuclear reduced mass, $V_g$ and $V_u$ are the BO potential curve calculated from the electronic structure of the $1\sigma_g$ and $2p\sigma_u$ states of $H_2^+$ and the coupling $V_{gu}=E(t)D_{gu}$ with $D_{gu}$ being the the electronic transition dipole between these two states. The electric field is expressed by
$
E_{\rm mix} = E_0 \exp [-2 \ln2 (\frac{t}{\tau_0})^2 ]\cos(\omega_0 t + \phi_{\rm CE})
+ E_1 \exp [-2 \ln2 (\frac{t}{\tau_1})^2 ]\cos(K \omega_0 t + \phi_{\rm CE} + \phi_1),
$
where $K$ is the frequency ratio between the main fundamental field ($\omega_0$)
and the weak mixing IR field ($\omega_1 = K \omega_0$,
$K<1$),
the subscript (0, 1) means the two laser field components, $E$ is
the electric field, $\tau$, $\phi_{\rm CE}$ and $\phi_1$ denote the pulse duration
(FWHM), the CEP and relative phase, respectively. Here $\phi_1=0$ and the CEPs of the fundamental and the IR signal lights are equal due to the phase relationship of OPA \citep{Baltuska}. The probability of electron localized on the left or right nucleus can be calculated by introducing the left and right localized wavefunction
$\psi_{l,r}=(\Psi_g\pm\Psi_u)\sqrt{2}$. The kinetic energy release (KER) spectrum $S_{l,r}(E_k)$ is calculated by Fourirer analysis of the real-space wavefunction with the bound states being projected out \cite{Abeln}.
The electron localization is quantified by the left-right
ejection asymmetry parameter $A(E_k,\phi_{\rm CEP})=\frac{S_l(E_k,\phi_{\rm CEP}) -S_r(E_k,\phi_{\rm CEP})}{S_l(E_k,\phi_{\rm CEP})+S_r(E_k,\phi_{\rm CEP})}$.

Figure \ref{fig1}(b) shows the electron localization asymmetry as a function of CEP $\phi_{\rm CE}$. The solid magenta and blue lines correspond to the interaction of SWS pulse with 4th and 8th vibrational states of $H_2^+$, respectively. For comparison, a similar simulation has been performed for the 5-fs OC field and the results are presented by the dashed red and green lines in Fig. \ref{fig1}(b). In both situations, we can observe sensitive CEP dependence of the electron localization. Nevertheless, in comparison with the 5-fs OC pulse, the asymmetry amplitude in the SWS pulse is two- and three-fold higher for the 8th and 4th vibrational states, respectively. Moreover, we need to emphasize that preparation mechanism of $H_2^+$ generally indicates an incoherent Franck-Condon(FC) distribution of vibrational states in experiments \cite{BenItzhak}.
Thus we have to average the observables over the initial vibrational states weighted by FC factors \cite{BenItzhak,Abeln}. 
Figure \ref{fig1}(c) shows the FC averaged localization asymmetry. However, we can see very distinct situations: in the OC field, the asymmetry has been significantly smeared. By contrast, very large asymmetry with amplitude of 0.35 still can be obtained in the SWS pulse, which corresponds to an electron
localization probability about $68\%$ of all fragmentation
events. We have changed the intensity from $2\times10^{13}$ to $2\times10^{14}$ W/cm$^2$ and pulse duration from 3 to 10 fs, the FC averaged asymmetry in the OC few-cycle pulse is below 0.06, which is only 1 sixth of that in the SWS pulse.
\begin{figure}[!htp]
\begin{center}
\includegraphics[width=8cm]{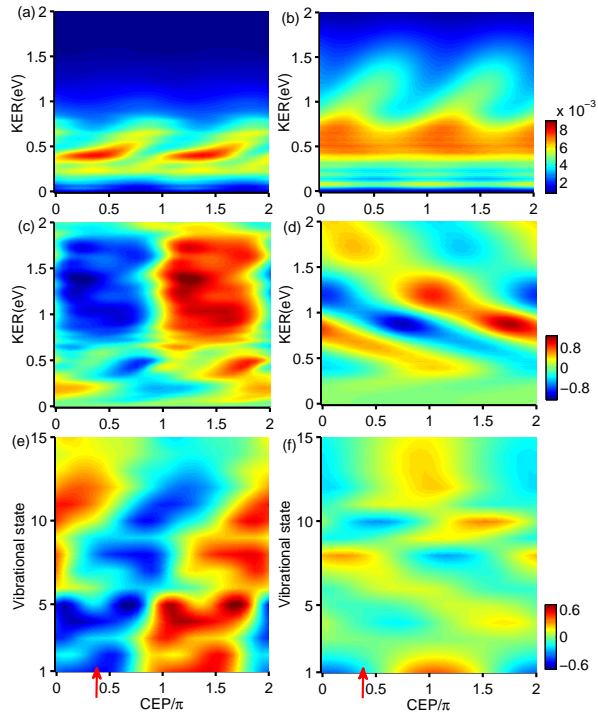}
\caption{\label{fig2}(Color) KER spectra as a function of CEP in the (a) SWS pulse (b) 5-fs pulse.
Asymmetry map as a function of KER and CEP in the (c) SWS pulse and (d) 5-fs pulse.
Asymmetry map as a function of vibrational state and CEP in the (e) SWS pulse and (f) 5-fs pulse.
The laser parameters are the same with Fig. \ref{fig1}(b).
}
\end{center}
\end{figure}

Figure \ref{fig2} shows the distinct difference of the electron localization between the SWS [(a), (c) and (e)] and OC 5-fs pulses [(b), (d) and (f)], respectively. From these results we can get several important remarks. (i) KER is lower in the SWS pulse and concentrated around 0.5 eV. The higher KER ($>1$ eV) is significantly suppressed compared with that of OC 5-fs pulse [see Fig. \ref{fig2}(a) and (b)]. (ii) The asymmetry electron localization depends sensitively on the CEP and KER in the OC 5-fs pulse, therefore resulting a ``tilted stripe'' as shown in Fig. \ref{fig2}(d). However, in the SWS field, the electron localization depends only sensitively on the CEP, but not sensitively on the KER. Therefore, as shown in Fig. \ref{fig2}(c), we observe a ``vertical stripe''. Note that the total population of the electron localization is the integration over the KER. Consequently, OC 5-fs totally results in a very weak contrast of the electron localization. Nevertheless SWS pulse can maintain the large contrast, resulting a much more effectively control of electron localization. (iii) The CEP dependence of electron localization is ``in-consistent-step'' for different vibrational states in the SWS field, however that is ``out-of-step'' in the OC field. As shown in Fig. \ref{fig2}(e), a ``tilted stripe'' can be observed in the asymmetry map as a function of CEP and vibrational states. Thus electron localization of some vibrational states (e.g., v=4 and 8 shown in Fig. \ref{fig1}(b)) are completely reverse, therefore smearing out the electron localization. However, in the SWS pulse, more electron is localized in the left nucleus in a broad range of CEP ($0.2\pi$ to $0.7\pi$) for the lowest 10 states. Note that the higher vibrational states correspond to a FC factor 1 order of magnitude lower, thus play a minor role. Such results clearly demonstrate the effective control of the electron localization with the SWS pulse. The FC-averaged asymmetry parameter is one order of magnitude higher than that of OC 5-fs scheme. Moreover, we should not that to keep the waveform of SWS pulse similar to a few-cycle pulse ($<5fs$), the pulse duration of the IR pulse should be less than 25 fs. Otherwise, different KER will show a different CEP dependence \cite{Cocke}, but the total electron localization asymmetry will be significantly washed out.

\begin{figure}[!htp]
\begin{center}
\includegraphics[width=\linewidth]{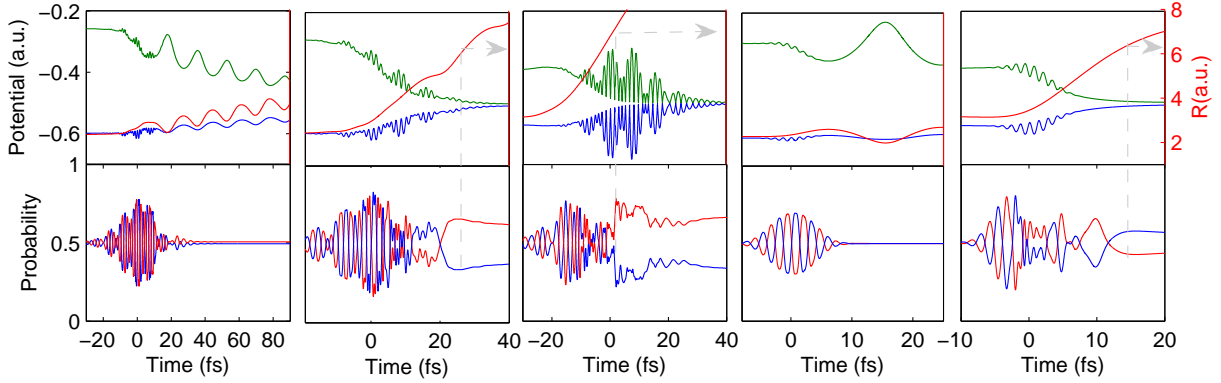}
\caption{\label{fig3} Color)Top plot: evolution of the eigenenergy of the quasi-states $\phi_1$ (green line) and $\phi_2$(blue line) and the internuclear distance (red line).
Bottom plot: evolution of the left (red line) and right (blue line) localized wavefunction $\mid\phi_{l,r}\mid^2$. (a), (b) and (c) correspond to the interaction of SWS pulse with the state v=3, 4 and 8, respectively. (d) and (e) correspond to the interaction of 5-fs pulse with the state v=3 and 8, respectively. The CEP $\phi_{\rm cep}=0.4\pi$ and other parameters are the same with Fig. \ref{fig1}.}
\end{center}
\end{figure}
In order to shed a deeper insight onto the difference of the electron localization in the SWS and OC few-cycle pulses schemes, we have analysis the molecular dynamics in terms of the quasi-static states,
which is related to left and right localized wave functions $\psi_{l,r}$ via \cite{Kono,quasiStatic},
$\psi_{1,2}=\frac{1}{\sqrt{2}}[\psi_{l}(\cos\theta\pm\sin\theta)\pm\psi_r(\cos\theta\mp\sin\theta)]$, where $2\theta=\tan^{-1}(-2V_{gu}/(V_u-V_g))$.
The corresponding quasi-static eigenvalues are $V_{1,2}=\frac{(V_g+V_u)}{2}\mp\sqrt{(V_u-V_g)^2/4+V_{gu}^2}$. For small internuclear distance $R<3$ au, the energy gap between $V_1$ and $V_2$ is much larger. Thus the coupling between $\phi_{1,2}$ is difficult, electron dynamics is adiabatic and can not be localized. As $R$ increase larger than 3.5 au, significant nonadiabatic coupling becomes possible and electron localization will be established. If $R>6$ au, the internuclear barrier is very high, which frozen the electron localization. Figure \ref{fig3} presents the evolution of the quasi-state eigenvalues $V_{1,2}$ and internuclear distance R (top plots) and the left and right localization probability (bottom plots) for different vibrational states subjected to the SWS and 5-fs OC pulses, respectively. Here the CEP $\phi_{\rm CE}$ is set to 0.4 $\pi$ [the corresponding asymmetry map is marked by the arrow in Fig. \ref{fig2}(e) and (f)]. For the interaction of lower vibrational state with 5-fs OC pulse, as shown in Fig. \ref{fig3}(d)(v=3), the motion of nuclear wavepacket is too slow to respect the ultrashort laser pulse. Thus the internuclear distance slightly changes around the equilibrium 2 au. The electron follows the oscillation of the laser field and changes its localization in each half cycle. The dynamics for higher vibrational states is completely different. As shown in Fig. \ref{fig3}(e)(v=8), the energy gap becomes smaller. The internuclear distance gradually increases larger than 6 au at the end of the pulse and more electron is localized near the left nucleus. By contrast, in the SWS pulse, the pedestal pulse is effective to stretch the internuclear distance, which increases to 4.5 au for V=3 and larger than 6 au for the higher states. Therefore, these vibration states show a similar nonadibiatic behavior. Of course, the CEP dependence can not be exactly same for different states, the different nevertheless becomes much smaller compared with that in the 5-fs pulse. This is possibly the reason why the CEP dependence is ``in consistent step'' for the SWS pulse. Moreover, because the internuclear distance is stretched, the dissociation happens at a large internuclear distance. Consequently, the KER is lower and the high KEP is significantly suppressed compared with that of the 5-fs pulse.

\begin{figure}[!htp]
\begin{center}
\includegraphics[width=12.0cm]{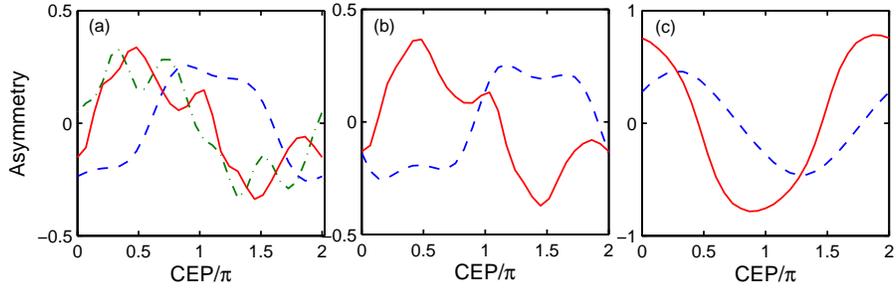}
\caption{\label{fig4}(Color) (a) CEP dependence of the asymmetry of electron localization of $H_2^+$ subjected to the pulse synthesized by 800-nm and 1300-nm (dashed blue line), 1500-nm (solid red line) and 1800-nm (dash-dotted green line), respectively. The pulse duration of the 800-nm pulse is 15 fs and is 25 fs for the other IR pulses. The laser intensity is the same with Fig. \ref{fig1}.
(b) Same to (a), but the target is changed to $D_2^+$ and the pulse duration of the 800-nm pulse is 20 fs and it 25 fs for the other IR pulse.
(c) Same to (a), but the target is changed to $H_2$. The pulse duration of the 800-nm pulse is 15 fs and is 25 fs for the other IR pulses. The laser intensity is the same with Fig. \ref{fig1}.
}
\end{center}
\end{figure}

Another advantage of our scheme of synthesis the coherent IR laser pulses output from OPA \cite{EJtak,Oliver} or OPCPA systems \citep{Kartner} is that the central wavelength of the ideal or signal lights can be adjusted in a broad range. This provides additional freedoms to control the sub-cycle waveform and the electron localization. Figure \ref{fig4}(a) shows the CEP dependence of asymmetry of electron localization by changing the IR wavelength to 1300 nm (blue dashed line), 1500 nm (red solid line) and 1800 nm (green dash-dotted line), respectively. A large asymmetry amplitude of about 0.3 can be obtained in all these cases, which corresponds to $65\%$ localization probability of the total dissociation events.
We have also simulated the electron localization for $D_2^+$. The CEP dependence of localization asymmetry is shown in Fig. \ref{fig4}(b) for the SWS field synthesized with 800-nm and 1200-nm (blue dashed line), 1500-nm (red solid line) IR pulses, respectively. The laser intensity is $1\times10^{14}$ $\rm W/cm^2$, relative intensity is $15\%$ and all the pulse durations are 25 fs. The localization asymmetry shown in Fig. \ref{fig4}(b) demonstrates the effectiveness of our SWS scheme in the case of $D_2^+$.

A different mechanism of electron localization lies on the recollision process \citep{Kling,KlingCO}. Briefly, this process starts from the ionization of deuteron molecules. The electron is accelerated in the laser field and recollides with the molecule ion as the electric field changes its direction. The recollision can excite the electron wavepacket and may localize the electron to one nucleus. Different from the previous process, where the electron localization is induced by the laser coupling between two states, the recollision process plays the vital role for electron localization. The SWS pulse also enables us to effectively control the recollision and so the electron localization. To demonstrate our scheme, we consider the target of $H_2$. As \citep{Kling}, we first we calculated the ionization of $H_2$ with the molecular Ammosov-Delone-Krainov (MOADK) ionization model \cite{MOADK}. The green filled area in Fig. \ref{fig1}(a) shows the ionization rate of $H_2$ subjected to the SWS pulse. One can see that the waveform of the SWS is similar to a 4-fs pulse. The electron mainly ionized at the highest peak, while the ionization rates at other peaks are less than $5\%$ of the highest peak. At each ionization time, the electron wavepacket of $H_2$ is launched to the $1s\sigma_u$ state of $H_2^+$. This wavepacket propagates for $0.7T_0$, then the electron recollides with the $H_2^+$ and boost the wavepacket to $2p\sigma_u$ state, which is repulsive and can be quickly dissociated. We use the same two-state model of TDSE to follow the motion of the wavefunction. Finally the resulting localization probability summed incoherently with the weight given by the ionization rate over all the wavepacket launched at different ionization times. Figure \ref{fig4}(c) shows the asymmetry of electron localization in the SWS pulse synthesized by 800-nm and 1200-nm (blue dashed line), 1500-nm (red solid line) IR pulses, respectively. One can see a very large asymmetry of electron localization, especially in the later case, the asymmetry amplitude is 0.78, which corresponds to an electron localization probability as high as $89\%$ of all fragmentation events. We should emphasize that the waveform of our SWS pulse is similar to a 4-fs pulse, which results in solely one ionization and recollision events. Such a character is crucial to effective control of electron localization, because the asymmetry maybe washed out if multi-recollision events contribute \cite{Kling}. Note also that Ref. \citep{Ullrich2} recently reported a relevant experiment, which demonstrate that, for molecule $H_2$, the electron localization induced by laser coupling also can be dominant. This process starts from the ionization of $H_2$ and then the wavepacket is coupled by the laser field but the recollision process plays a minor role. We have also simulated this process with the SWS pulse. Our results have indicates a high degree of control over the electron localization (asymmetry parameter is about $0.4$).

In conclusion, we have shown an effective scheme based on the SWS by synthesizing multicycle IR pulse of different wavelengthes to control the electron localization of $H_2^+$, $H_2$ and its isotopes. Compared with the few-cycle pulse, SWS pulse shows a higher degree of electron localization control. In the case of $H_2^+$, a localization probability of $68\%$ are realized and for $H_2$, the localization possibility can be as high as $89\%$ with the SWS scheme. Moreover, the current laser technology has made it possible to synthesize and stabilize the phase of multicolor coherent pulses, e.g., the CEP-stabilized pump, signal, idle lights of optical parametric synthesizers. We may expect an effective control of the electron localization for more complex molecules than hydrogen by multicolor SWS.

\end{document}